\documentclass[12pt,preprint]{aastex}

\slugcomment{Submitted to Astrophysical Journal}

\shorttitle{Solar flare Cl abundance}
\shortauthors{Sylwester et al.}

\begin{document}

\title{The Solar Flare Chlorine Abundance from RESIK X-ray Spectra}


\author{B. Sylwester }
\affil{Space Research Centre, Polish Academy of Sciences, 51-622, Kopernika~11, Wroc{\l}aw, Poland}
\email{bs@cbk.pan.wroc.pl}

\author{K. J. H. Phillips}
\affil{Mullard Space Science Laboratory, University College London, Holmbury St Mary, Dorking,
Surrey RH5 6NT, U.K.}
\email{kjhp@mssl.ucl.ac.uk}

\author{J. Sylwester }
\affil{Space Research Centre, Polish Academy of Sciences, 51-622, Kopernika~11, Wroc{\l}aw, Poland}
\email{js@cbk.pan.wroc.pl}

\and

\author{V. D. Kuznetsov }
\affil{Institute of Terrestrial Magnetism and Radiowave Propagation (IZMIRAN), Troitsk, Moscow, Russia}
\email{kvd@izmiran.ru}

\begin{abstract}
The abundance of chlorine is determined from X-ray spectra obtained with the RESIK instrument on {\em CORONAS-F} during solar flares between 2002 and 2003. Using weak lines of He-like Cl, \ion{Cl}{16}, between 4.44 and 4.50~\AA, and with temperatures and emission measures from {\em GOES} on an isothermal assumption, we obtained $A({\rm Cl}) = 5.75 \pm 0.26$ on a scale $A({\rm H}) = 12$. The uncertainty reflects an approximately factor 2 scatter in measured line fluxes. Nevertheless our value represents what is probably the best solar determination yet obtained. It is higher by factors of 1.8 and 2.7 than Cl abundance estimates from an infrared sunspot spectrum and nearby \ion{H}{2} regions. The constancy of the RESIK abundance values over a large range of flares ({\em GOES} class from below C1 to X1) argues for any fractionation that may be present in the low solar atmosphere to be independent of the degree of solar activity.
\end{abstract}


\section{INTRODUCTION}\label{intro}

Chlorine is an odd-$Z$ element with low abundance in the solar photosphere and in meteorites. Chlorine has no photospheric lines in the visible spectrum available for abundance analysis, though an infrared H$^{35}$Cl sunspot spectrum of \cite{hal72} gave $A(^{35}{\rm Cl}) = 5.4 \pm 0.3$ (abundances expressed on a logarithmic scale with $A({\rm H}) = 12$). Correcting this for the $^{37}$Cl/$^{35}$Cl isotope ratio (= 1/3) gives $A({\rm Cl}) = 5.5 \pm 0.3$. An upper limit of $A({\rm Cl})= 5.5$ has been deduced by \cite{lam71} from the near-infrared (8375~\AA) photospheric line. \cite{asp09} comment that the Cl abundance estimate of $A({\rm Cl}) = 5.32 \pm 0.07$ \citep{gar07} from \ion{H}{2} regions may be more reliable as a solar abundance than direct solar measurements. Chlorine abundances from interstellar medium lines to nearby stars range from 5.5 to 5.7 \citep{yor83,kee86,har84}. \ion{Cl}{1} lines occur in the solar ultraviolet spectrum, with that at 1351.7~\AA\ anomalously strong through pumping by the \ion{C}{2} 1335~\AA\ line \citep{shi83}, while \cite{fel04} have identified Cl in ionization stages from \ion{Cl}{10} to \ion{Cl}{12} in SUMER spectra. The very low intensities of these lines are unlikely to lead to a reliable determination of the quiet-Sun corona Cl abundance.  H-like and He-like Cl  (\ion{Cl}{17} and \ion{Cl}{16}) X-ray emission lines occur in solar flare spectra near 4~\AA. A scan of the {\em Solar Maximum Mission} Flat Crystal Spectrometer (FCS) over a very weak feature identified as the  \ion{Cl}{16} resonance line (4.444~\AA) and a nearby \ion{S}{15} line at 4.299~\AA\ during a powerful (class X1.4) flare in 1988 (discussed by \cite{phi90}) gave the Cl/S abundance ratio equal to $13.5 \pm 40$~\%, or (with \cite{fel00}'s S abundance) $A({\rm Cl}) = 6.1 \pm 0.15$. This is significantly higher than the estimates of \cite{hal72} or \cite{gar07}.

This paper reports on the observation of \ion{Cl}{16} lines with the RESIK (REntgenovsky Spektrometr s Izognutymi Kristalami; \cite{syl05}) crystal spectrometer on the {\em CORONAS-F} spacecraft during 20 solar flares, from which it has been possible to determine much more definitively the Cl abundance for flare plasmas. The detection of the lines has already been reported in an earlier version of the present work \citep{syl04} based on a smaller sample of RESIK data. The \ion{Cl}{16} resonance line and nearby intercombination and forbidden lines are weak but with the much higher sensitivity of RESIK than the FCS and other flat-crystal instruments the lines can be distinguished fairly well in flare spectra over a large temperature range, and from the measured line fluxes together with the assumption of an isothermal plasma estimates of the chlorine abundance made. These are compared with those of other authors, and the relevance to the widely discussed first ionization potential (FIP) dependence of coronal abundances discussed.

\section{RESIK FLARE SPECTRA}

The instrumental details of RESIK, which operated for nearly 2 years from spacecraft launch on 2001 July~31, have been given elsewhere \citep{syl05}. In summary, RESIK had combinations of four curved crystals and position-sensitive proportional counter arranged in pairs. There were no collimators in order to maximize the instrument's sensitivity. Photon counts in each spectral range were accumulated in data gathering intervals (DGI) that were fixed at 10~s for the initial few months of the mission, but for most of the period analyzed here the DGIs varied from 2~s at the peaks of strong flares to about 5~minutes when activity was low. The total spectral range was 3.3--6.1~\AA. The \ion{Cl}{16} lines between 4.444~\AA\ and 4.497~\AA\ discussed here occur within RESIK's channel~3 (range 4.32--4.86~\AA). Channel~3 also includes the \ion{S}{16} Ly-$\alpha$ line at 4.729~\AA\ which features strongly at $T_e \gtrsim 10$~MK. The diffracting crystal for channel~3 was quartz ($10{\bar1}0$), and the spectral resolution 12~m\AA. Post-launch sensitivity calculations, discussed at length by \cite{syl05}, enable absolute fluxes of spectral lines to be determined to $\sim 20$\% precision. A background flux due to fluorescence was found from 32-bin pulse height analyzer data, from which peaks due to solar and fluorescence counts can be distinguished. By fitting gaussian distributions to the peaks, the fluorescence background was found to be negligible for channels~1 and 2 and approximately estimated and subtracted from spectra in channels~3 and 4.

In previous analyses of RESIK Ar and K line spectra and continua emitted at flare temperatures \citep{syl10b,syl10c,phi10}, an isothermal emitting plasma with temperature and emission measure given by {\em GOES} emission was assumed and found to be valid. However, this assumption ceases to be a good approximation for lower-temperature emission, e.g. that observed by RESIK in quiet Sun conditions, for which a differential emission measure technique was used instead \citep{syl10a}. As the contribution functions $G(T_e)$, defining the emission per unit emission measure $N_e^2 V$ of the emitting plasma ($N_e = $ electron density, $V$ the emitting volume) of \ion{Cl}{16} lines peak at $T_e \sim 15$~MK, comparable to the corresponding temperatures for \ion{Ar}{17}, \ion{Ar}{18}, or \ion{K}{18} lines, an isothermal approximation was again used, with temperatures (called here $T_{\rm GOES}$) and emission measures ($EM_{\rm GOES}$) taken from the ratio of the emission in the two X-ray channels of {\em GOES} using a coronal set of abundances \citep{fel92}. It was not possible to use \ion{Cl}{15} satellite line ratios for the temperature as at the RESIK spectral resolution the satellites are blended with the \ion{Cl}{16} lines. As was discussed by \cite{syl10b}, \cite{syl10c}, and \cite{phi10} and will be discussed further here, use of the {\em GOES} ratios appears to be a valid assumption for our purposes.

\begin{figure}
\epsscale{.80}
\plotone{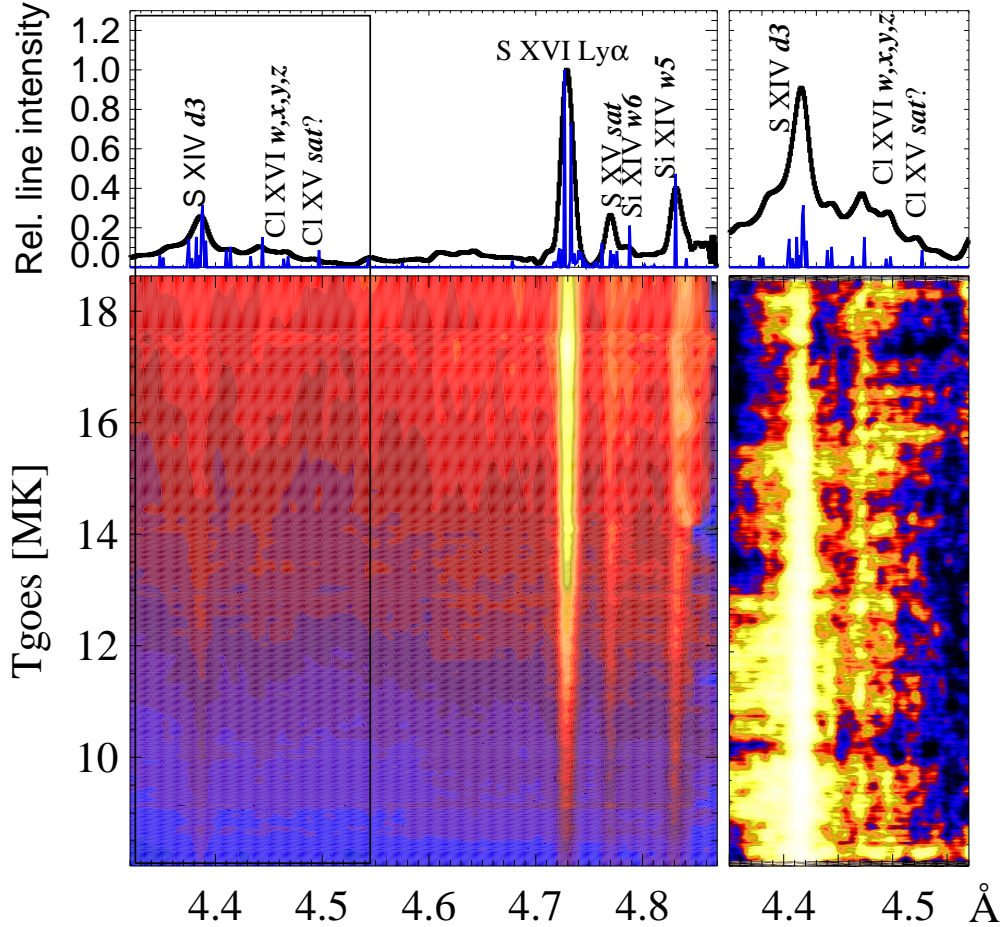}
\caption{Left: RESIK channel~3 spectra (4.32--4.87~\AA) during flares stacked vertically with temperature from {\em GOES} ($T_{\rm GOES}$) increasing upwards (scale indicated) and (top) the total spectrum (dark continuous curve) and matching synthesized spectrum, with principal lines identified.  Right: Portion of RESIK channel~3 spectra, wavelength range 4.35--4.57~\AA, indicated in the box of the left panel, showing the \ion{S}{14} satellite feature at 4.39~\AA\ and the \ion{Cl}{16} lines between 4.44~\AA\ and 4.50~\AA. To show the Cl lines more clearly, a square-root intensity scale has been used. As with the left panel, $T_{\rm GOES}$ increases upwards. (A color version of this figure is available in the online journal.)  } \label{chan3_stacked_spectra}
\end{figure}

A total of 2795 RESIK spectra were analyzed, taken during 20 flares from 2002 August~3 to 2003 February~22. A list of the observations was given by \cite{phi10}. A convenient display of the entire set of observed spectra obtained during the {\em CORONAS-F} mission lifetime is formed by stacking the spectra on a common wavelength scale, arranging them in order of $T_{\rm GOES}$ determined at the midpoint time over which each spectrum was accumulated, colors or a gray-scale indicating the intensity of the emission. Such displays were given by \cite{syl10b} and \cite{syl10c} for channels~1 and 2 respectively. Figure~\ref{chan3_stacked_spectra} (left panel) shows in the same manner all RESIK spectra in channel~3 which includes the \ion{Cl}{16} lines. An averaged spectrum in this range over the entire period is shown above the stacked spectra, with the principal spectral lines indicated. Owing to the low abundance of Cl, the \ion{Cl}{16} lines are very weak, so a portion of the display around their wavelengths and the feature at 4.39~\AA\ made up of \ion{S}{14} dielectronic satellite lines (transitions $1s^2 nl - 1s 3p nl$) is shown in the right panel of Figure~\ref{chan3_stacked_spectra}, with an intensity scaling that shows the \ion{Cl}{16} lines more clearly. Details of the \ion{Cl}{16} lines are given in Table~\ref{Cl_line_data}.

\begin{figure}
\epsscale{1.0}
\plottwo{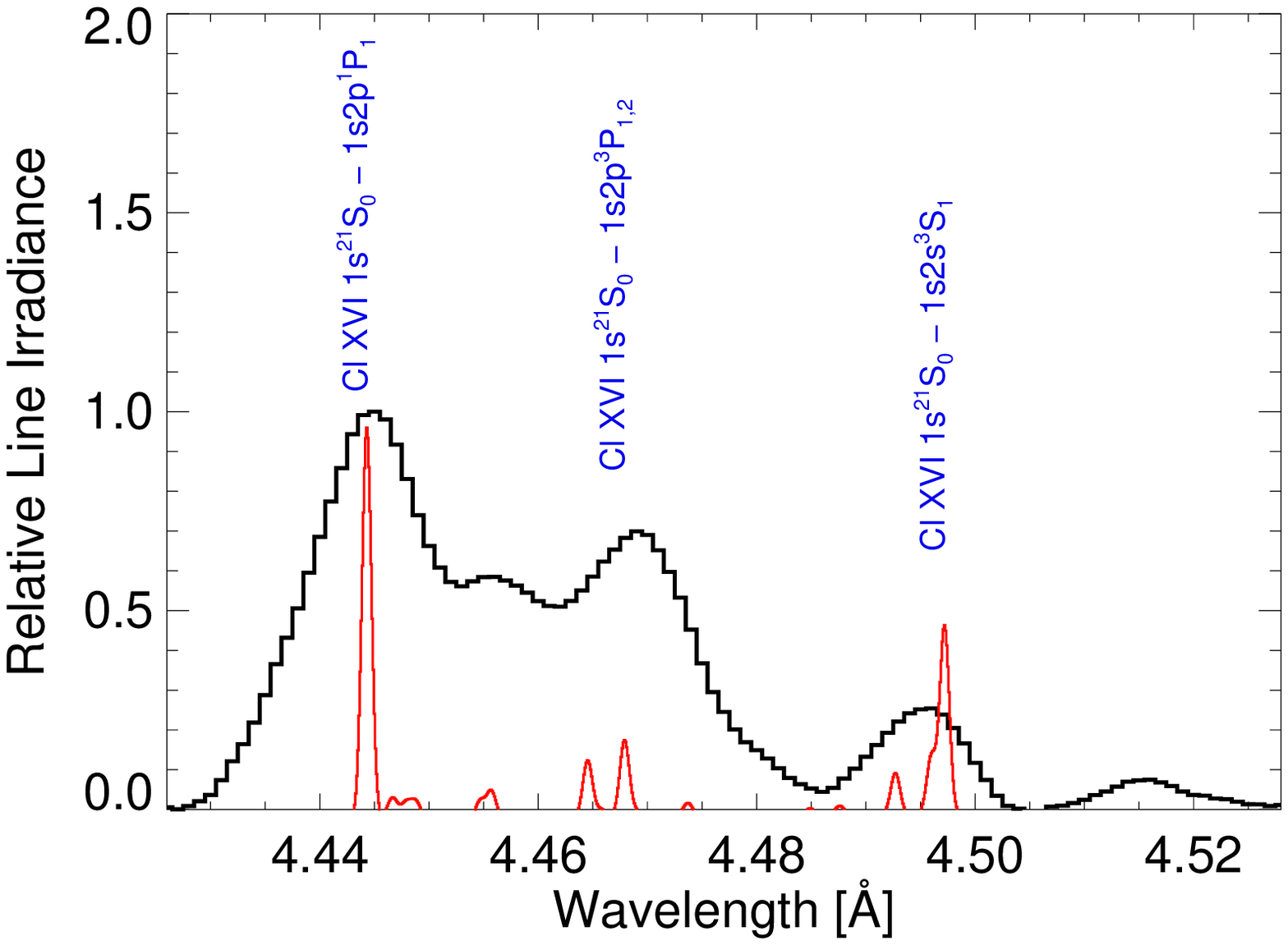}{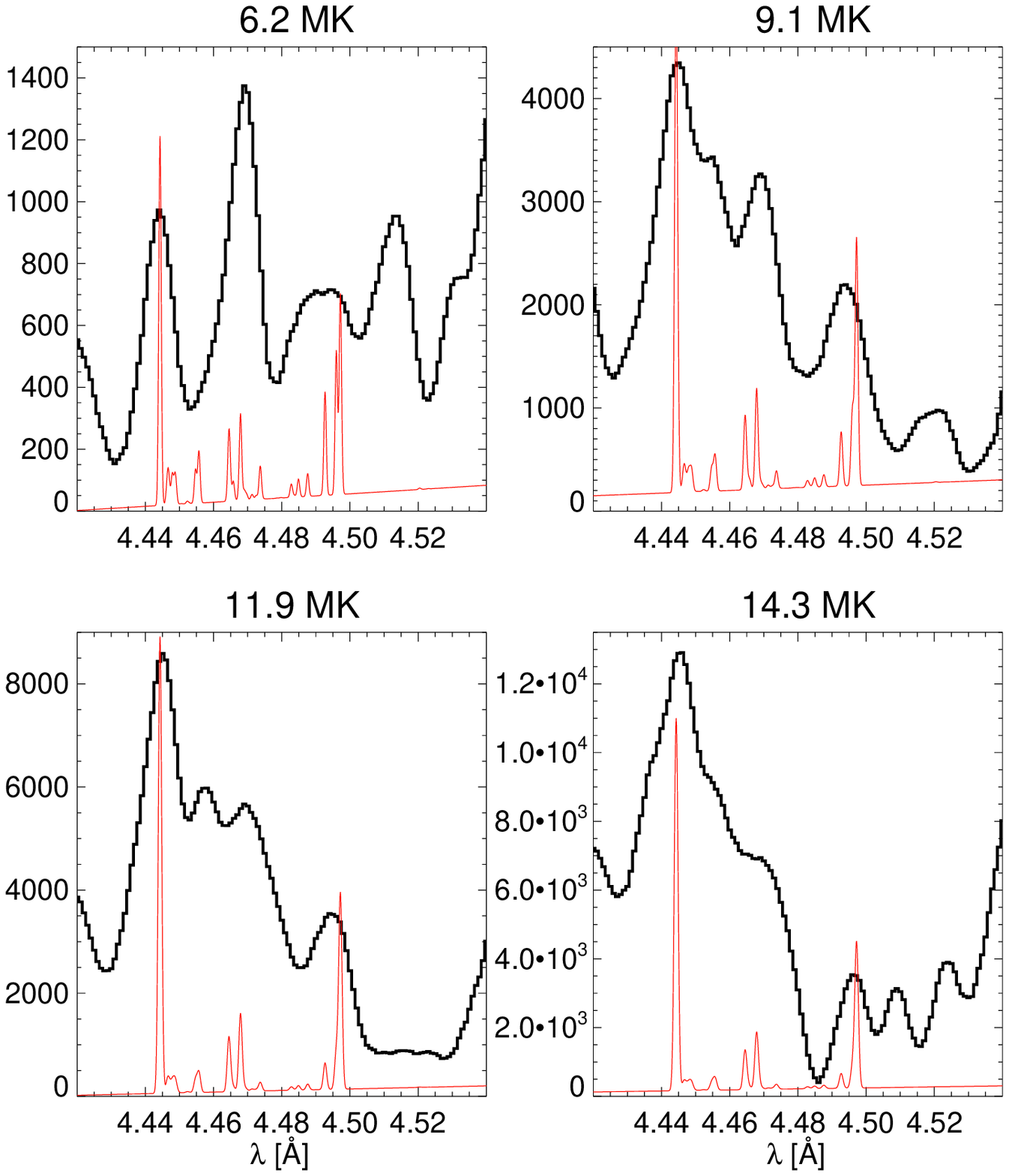}
\caption{(Left panel) Summed RESIK channel~3 spectra in the 4.425--4.53~\AA\ range for all flares, with a theoretical spectrum (red curve) calculated for a temperature of 9~MK, showing the positions of the principal \ion{Cl}{16} lines; (Right panel) RESIK channel~3 spectra (4.42--4.54~\AA) showing the \ion{Cl}{16} lines (black curves) with synthesized spectra (red curves). RESIK spectra are averaged over several spectra in narrow ranges of $T_{\rm GOES}$ (the mean of the range is indicated in MK above each spectrum) and normalized to an emission measure of $10^{48}$~cm$^{-3}$. The vertical (spectral flux) scale is in units of photons cm$^{-2}$ s$^{-1}$ \AA$^{-1}$. (A color version of this figure is available in the online journal.) } \label{ClXVI_spectra}
\end{figure}


\section{SYNTHETIC SPECTRA}\label{synth_sp}

A computer program was written to synthesize the spectral region 4.38--4.58~\AA\ containing the \ion{Cl}{16} lines as a function of temperature $T_e$ and emission measure $N_e^2 V$ so that comparison could be made with RESIK spectra at particular values of the observed temperature $T_{\rm GOES}$. We used the {\sc chianti} database for the line fluxes of the \ion{Cl}{16} lines. The CHIANTI collisional excitation data for the lines are interpolated using data from various sources for He-like ions with atomic numbers similar to Cl. Data for \ion{Cl}{15} dielectronic satellites, the fluxes of which are important at temperatures $T_e \lesssim 13$~MK, are not included in {\sc chianti}. The satellites of interest have transitions $1s^2 nl - 1s2p\,nl$ ($nl$ representing the non-participating or spectator electron), and are excited by dielectronic recombination of the He-like ion Cl$^{+15}$ and, for satellites in the $1s^2 2s - 1s2s2p$ array, by inner-shell excitation of the Li-like ion Cl$^{+14}$. Only a few dielectronically formed satellites are important individually, but calculations need to be done for a fairly large range of satellites as the cumulative effect of satellites with high values of $n$ is significant in adding to the intensities of \ion{Cl}{16} lines $w$ and $y$. The flux of a dielectronically formed satellite emitted by a plasma with temperature $T_e$ and emission measure $N_e^2 V$ at the distance of the Earth (1~AU) is given by

\begin{equation}
I(s) = 1.66 \times 10^{-16} \,\,\frac{N({\rm Cl}^{+15})}{N({\rm Cl})} \frac{N({\rm Cl})}{N({\rm H})} \frac{N({\rm H})}{N_e} \frac{N_e^2 V}{4\pi ({\rm AU})^2} \frac{F(s)\, {\rm exp}\,(-\Delta E/kT_e)}{T_e^{3/2}} \,\,\,{\rm photon}\,\, {\rm cm}^{-2} \,\,{\rm s}^{-1}
\end{equation}\label{sat_line}

\noindent where the ion fractions $N({\rm Cl}^{+15})/N({\rm Cl})$ as a function of $T_e$ were taken from \cite{bry09}; these are negligibly different from those of \cite{der09} which are the default set in {\sc chianti}. The value of $N({\rm H})/N_e = 0.8$ for a flare plasma. The satellite intensity factor $F(s)$  is a function of autoionization and radiative rates from the upper level, and the excitation energy $\Delta E$ is with respect to the ground level of the He-like ion.

The satellite wavelengths and $F(s)$ values were calculated from the \cite{cow81}  Hartree--Fock atomic code with pseudo-relativistic corrections (HFR), adapted for small personal computers (A. Kramida 2008, private communication). A total of 207 transitions were considered with spectator electrons with $nl$ up to $4f$. As in previous runs of this code (e.g. \cite{syl10b}), we used 100\% scaling factors for the Slater parameters in the HFR code. The satellite wavelengths generally need small adjustments to bring them into agreement with observed values. The adjustments (of $+3$~m\AA) were derived from the fact that high-$n$ satellites converge on the \ion{Cl}{16} $w$ and $y$ lines. Values of $F(s)$, $\Delta E$, and wavelengths for the most intense satellites (defined to be those with $F(s) > 4 \times 10^{13}$~s$^{-1}$) are given in Table~\ref{Cl_line_data}.

\begin{deluxetable}{lcccc}
\tabletypesize{\scriptsize} \tablecaption{P{\sc rincipal} Cl L{\sc ines} {\sc in the} 4.45--4.50 \AA\ R{\sc ange} \label{Cl_line_data}} \tablewidth{0pt}

\tablehead{\colhead{Transition} & \colhead{Wavelength (\AA)} & \colhead{Notation$^a$} & \colhead{$F(s)$ (s$^{-1}$)} & \colhead{$\Delta E$ (keV) }}

\startdata
{\em Cl XVI lines}\\

$1s^2\,^1S_0 - 1s2p\,^1P_1$ & 4.444 &  $w$ \\
$1s^2\,^1S_0 - 1s2p\,^3P_2$ & 4.464 &  $x$ \\
$1s^2\,^1S_0 - 1s2p\,^3P_1$ & 4.468 &  $y$ \\
$1s^2\,^1S_0 - 1s2s\,^3S_1$ & 4.497 &  $z$ \\
\\

{\em Cl XV satellites}\\
$1s^2 4d\,\,^2P_{3/2} - 1s\,2p\,4d\,\,(^1P)\,^2F_{7/2}$  & 4.447 & &$5.12 (13)$ &  2.59  \\
$1s^2 3d\,\,^2D_{3/2} - 1s\,2p\,3d\,\,(^1P)\,^2F_{5/2}$  & 4.447 & &$5.35 (13)$ &  2.44  \\
$1s^2 3d\,\,^2D_{5/2} - 1s\,2p\,3d\,\,(^1P)\,^2F_{7/2}$  & 4.448 & &$8.80 (13)$ &  2.44  \\
$1s^2 4p\,\,^2P_{1/2} - 1s\,2p\,4p\,\,(^1P)\,^2D_{3/2}$  & 4.449 & &$4.04 (13)$ &  2.59 \\
$1s^2 4p\,\,^2P_{3/2} - 1s\,2p\,4p\,\,(^1P)\,^2D_{5/2}$  & 4.449 & &$5.35 (13)$ &  2.59  \\
$1s^2 3p\,\,^2P_{1/2} - 1s\,2p\,3p\,\,(^1P)\,^2D_{3/2}$  & 4.455 & $d15 $ &$1.07 (14)$ &  2.43  \\
$1s^2 3p\,\,^2P_{3/2} - 1s\,2p\,3p\,\,(^1P)\,^2D_{5/2}$  & 4.456 & $d13 $ &$1.60 (14)$ &  2.43  \\
$1s^2 2p\,\,^2P_{3/2} - 1s\,2p^2\,\,(^1S)\,^2S_{1/2}$  & 4.474 & $m$ & $4.53 (13)$ &  1.98  \\
$1s^2 2p\,\,^2P_{1/2} - 1s\,2p^2\,\,(^1D)\,^2D_{3/2}$  & 4.493 & $k$ & $1.42 (14)$ &  1.98  \\
$1s^2 2p\,\,^2P_{3/2} - 1s\,2p^2\,\,(^1D)\,^2D_{5/2}$  & 4.496 & $j$ & $1.97 (14)$ &  1.98 \\

\\
\enddata

\tablenotetext{a}{Notation of \cite{gab72, bel79}. The Cl~XVI wavelengths are from \cite{kel87}; the Cl~XV wavelengths are from the Cowan HFR code + 0.003~\AA. }
\end{deluxetable}

For the $1s^2 2s - 1s2s2p$ satellites we used the collision strength data of \cite{bel82} for \ion{Ca}{18} lines to calculate the inner-shell excitation rates. Because of the lower atomic number of Cl, these rates are probably a little higher than those for \ion{Cl}{15}, but the effect of the error on the final spectral calculation is likely to be negligible.

The neighboring free--free and free--bound continua were calculated from  {\sc chianti} routines. The free--bound continuum calculation requires a set of abundances, notably Si, Fe, Mg, and O. In an analysis of four X-ray continuum bands observed by RESIK \citep{phi10}, it was found that a coronal set of abundances (\cite{fel92}, given in the {\sc chianti} database  as the file sun$\_$corona$\_$ext.abund) fitted the RESIK continuum fluxes better (by a factor 2) than photospheric abundances.  We therefore used this coronal set in the present analysis. There is a slight danger here of a circular argument since the \ion{Cl}{16} spectrum will be used (Section~\ref{abundance_of_cl}) to derive the Cl abundance, and the result could be affected if Cl makes a significant contribution to the free--bound continuum. However, the {\sc chianti} calculations show that this contribution is negligible.

This synthetic spectrum program was then used to calculate spectra with temperature and emission measure as input. These are compared with RESIK spectra in Figure~\ref{ClXVI_spectra}. The left panel shows the sum of all RESIK channel~3 spectra taken over all flares observed, and shows that, despite the weakness of the \ion{Cl}{16} emission, the three main line features are evident. The right panel shows four RESIK channel~3 spectra in the 4.42--4.54~\AA\ range, including the \ion{Cl}{16} lines, for different values of $T_{\rm GOES}$ (indicated in each plot). Calculated spectra from the spectral synthesis program are shown in each case (by red continuous lines in the color version of the Journal), with temperatures equal to 9~MK for the summed spectra (left panel) and equal to values of $T_{\rm GOES}$ (right panel) for the others. The line profiles of the calculated spectra are defined by thermal Doppler broadening appropriate to each temperature (e.g. FWHM $= 1.6$~m\AA\ for a temperature of 9~MK). As can be seen, the weakness of the \ion{Cl}{16} lines makes the comparison somewhat difficult, particularly at lower temperatures, but there is no doubt that the chief features corresponding to the $w$, $x+y$, and $z$ lines are present.

The synthetic spectrum program was also used to reproduce the flare spectrum from the {\em SMM} Flat Crystal Spectrometer discussed by \cite{phi90} using the {\em GOES} temperature and emission measure at the time when the \ion{Cl}{16} lines were scanned. The theoretical and observed \ion{Cl}{16} $w$ line flux agree to $\sim 20$~\%. This confirms our previous analyses of RESIK spectra that, in lieu of more precise temperature-determining means such as dielectronic satellite ratios (satellites are not resolved at the resolution of RESIK), {\em GOES} temperatures and emission measures accurately represent the spectra discussed here.

\section{ABUNDANCE OF Cl FROM RESIK}\label{abundance_of_cl}

\begin{figure}
\epsscale{.80}
\plotone{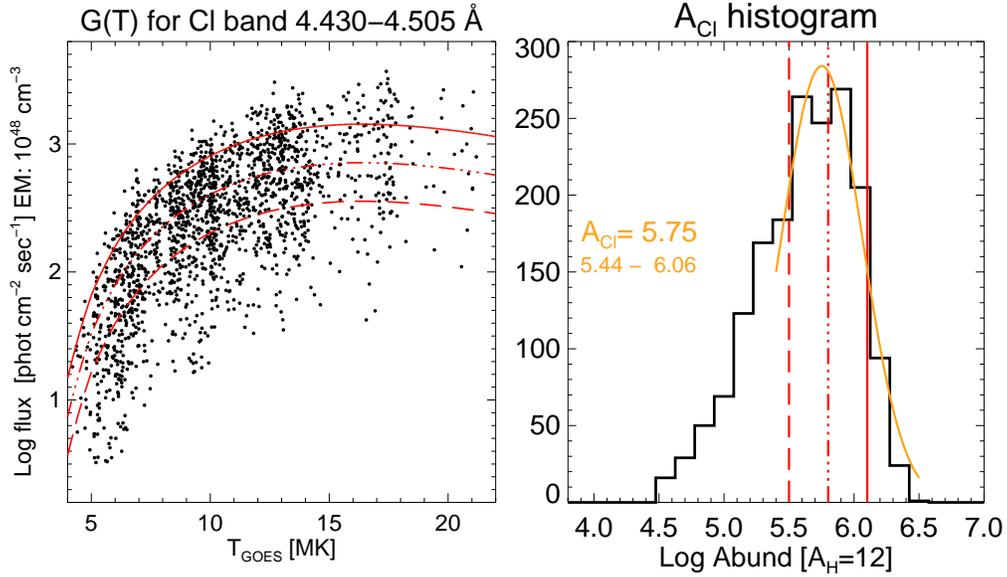}
\caption{(Left panel:) Fluxes of the total \ion{Cl}{16} line emission (including \ion{Cl}{15} satellites) in the 4.43--4.505~\AA\ divided by emission measures from {\it GOES} ($EM_{\rm GOES}$, units of $10^{48}$~cm$^{-3}$) plotted against $T_{\rm GOES}$ (black points). These are compared with the theoretical $G(T_e)$ curves for the total line emission assuming $A({\rm Cl})=5.5$ \citep{hal72}, shown as the lowest dashed curve, and 2 and 4 times this abundance (middle dot--dash curve and top solid curve). (Right panel:) Distribution of Cl abundance $A({\rm Cl})$ estimated from the \ion{Cl}{16} line emission (histogram). The peak of the best-fitting Gaussian distribution is $A({\rm Cl}) = 5.75$, with width (FWHM) range equal to 5.44--6.06 (corresponding to a $1 \sigma$ uncertainty of 0.26). The vertical lines indicate the value $A({\rm Cl}) = 5.5$ from \cite{hal72} and 2 and 4 times this abundance. (A color version of this figure is available in the on-line journal.) } \label{G_of_T_Cl_XVI}
\end{figure}

Following procedures used for K and Ar \citep{syl10b,syl10c}, the observed \ion{Cl}{16} line emission for all spectra were compared with the summed contribution functions $G(T_e)$ for these lines. For each spectrum we measured the total emission $F_i$ (in photons cm$^{-2}$~s$^{-1}$) above a neighboring pedestal level in the 4.43--4.505~\AA\ band, which includes the \ion{Cl}{16} $w$, $x+y$, and $z$ lines and nearby \ion{Cl}{15} satellites. A plot of the values of $F_i/EM_{\rm GOES}$ (where $EM_{\rm GOES}$ is in units of $10^{48}$~cm$^{-3}$) against $T_{\rm GOES}$ for all spectra is shown as points in Figure~\ref{G_of_T_Cl_XVI} (left panel) together with the theoretical $G(T_e)$  (continuous curve), calculated from the synthetic spectrum program using $A({\rm Cl}) = 5.5$ \citep{hal72} and 2 and 4 times this value. The total scatter in the points covers a range of nearly a factor 30, reflecting the weakness of the \ion{Cl}{16} lines in RESIK spectra over the 4--21~MK temperature range that the lines are discernible. The scatter is comparable to our results from similarly weak \ion{K}{18} lines \citep{syl10b} in RESIK channel~1 but is greater than those from the much stronger \ion{Ar}{17} lines in channel~2 \citep{syl10c}. It is clear, nevertheless, that the cluster of points, while following the trend with temperature of the theoretical curve (steep rise at low temperatures, maximizing at $\sim 17$~MK), mostly fall above it. A chlorine abundance greater than that given by \cite{hal72} is therefore indicated.

The abundance of Cl can be estimated from each RESIK spectrum using Figure~\ref{G_of_T_Cl_XVI} (left panel) by taking the abundance to be $f_i A_{\rm HN}({\rm Cl})$ where $A_{\rm HN}({\rm Cl})$ is the Cl abundance of \cite{hal72} and the correcting factor $f_i$ for the $i$th spectrum is evaluated from

\begin{equation}
f_i = \frac{F_i}{G(T_{\rm GOES}) EM_{\rm GOES}}
\end{equation}

\noindent where $F_i$ is the total flux of the \ion{Cl}{16} lines and \ion{Cl}{15} satellites and the contribution function $G$ is evaluated for temperature $T_{\rm GOES}$ and the Cl abundance of \cite{hal72}. The distribution of all $A({\rm Cl}) = f A_{\rm HN}({\rm Cl})$ measured values in bins of 0.1 in the logarithm of abundance is shown in Figure~\ref{G_of_T_Cl_XVI} (right panel). The peak of a best-fit  Gaussian distribution is at $A({\rm Cl})=5.75$, with width (FWHM) range 5.44--6.06. The corresponding $1\sigma$ uncertainty is 0.26.


\section{SUMMARY AND CONCLUSIONS}

The large number of RESIK channel~3 spectra obtained over the lifetime of the \newline
{\em CORONAS~-~F} mission has enabled the abundance of chlorine to be determined from the \ion{Cl}{16} lines in the 4.44--4.50~\AA\ range. The value from the \ion{Cl}{16} lines is $A({\rm Cl})=5.75 \pm 0.26$, the uncertainty being based on the scatter of the 2795 observational points. This is a factor of 1.8 higher than the abundance determined from the infrared sunspot spectrum by \cite{hal72} and a factor 2.7 higher than that from \ion{H}{2} regions \citep{gar07} which \cite{asp09} consider may be a solar Cl abundance proxy.

According to \cite{fel00}, elements with low ($<10$~eV) FIP appear to have coronal abundances that are enhanced over photospheric abundances by a factor $\sim 4$, the so-called FIP effect. The first ionization potential of Cl is 12.97~eV \citep{all73}, so is by this definition a high-FIP element and is not expected to be enhanced in the corona. However, the value obtained here from a sample of flares appears to be a factor of almost 2 higher than the sunspot abundance of \cite{hal72}, against expectation of the FIP effect for other elements. Although abundance determinations from individual \ion{Cl}{16} spectra have some scatter (Figure~\ref{G_of_T_Cl_XVI}, left panel), the distribution of the determinations (Figure~\ref{G_of_T_Cl_XVI}, right panel) has a width sufficiently small to suggest consistency  arguing for the constancy of $A({\rm Cl})$ with time. This was also found to be the case, based on the same data sample, for K (with similar uncertainties to those for Cl: \cite{syl10b}) but more particularly for Ar, having only $\sim 20$\% uncertainties \citep{syl10c}. If indeed there is a fractionation process that separates ions from neutrals at some low level in the solar atmosphere giving rise to the FIP effect, the process appears to be a steady one and not particularly dependent on the occurrence of flares or of their  importance: the flares used in the analysis here range in importance from below C1 to X1 (see list in \cite{phi10}), a factor of over 100 in soft X-ray output.

\acknowledgments
We acknowledge financial support from the European Commission's Seventh Framework Programme (FP7/2007--2013) under grant agreement No. 218816 (SOTERIA project, www.soteria-space.eu), the Polish Ministry of Education and Science Grant N N203 381736,  and the UK--Royal Society/Polish Academy of Sciences International Joint Project (grant number 2006/R3) for travel support. {\sc chianti} is a collaborative project involving Naval Research Laboratory (USA), the Universities of Florence (Italy) and Cambridge (UK), and George Mason University (USA).  We thank A. Kramida for help with running the Cowan HFR code in the form for small personal computers.


\end{document}